\documentstyle[psfig,12pt]{article}
\begin{document}
\setlength{\textheight}{220mm}
\setlength{\textwidth}{160mm}
\topmargin=-2cm

\title{Antikaon production in nucleon-nucleon reactions near 
threshold\footnote{Supported by BMBF, GSI Darmstadt and 
Forschungszentrum J\"ulich GmbH}}
\bigskip
\author{A. Sibirtsev, W. Cassing, and C. M. Ko\thanks{Permanent address:
Cyclotron Institute and Physics Department, Texas A\&M University,
College Station, Texas 77843, USA} \\
Institut f\"ur Theoretische Physik, Universit\"at Giessen \\
D-35392 Giessen, Germany}
\date{}
\maketitle

\begin{abstract}

The antikaon production cross section from nucleon-nucleon reactions
near threshold is studied in a meson exchange model. We include
both pion and kaon exchange, but neglect
the interference between the amplitudes.
In case of pion exchange the
antikaon production cross section can be expressed in terms of 
the antikaon production cross section from a pion-nucleon interaction,
which we take from the experimental data if available. Otherwise,
a $K^*$-resonance exchange model is introduced to relate the different
reaction cross sections.
In case of kaon exchange the antikaon production
cross section is related to the elastic $KN$ and $\bar KN$ cross sections,
which are again taken from experimental measurements.  We find that
the one-meson exchange model gives a satisfactory fit 
to the available data for the $NN\to NNK\bar K$ cross section at high 
energies.  We compare our predictions for the cross section near threshold 
with an earlier empirical parameterization and that from phase space models.
\end{abstract}

\newpage

\section{Introduction}

All data on antikaon production cross sections in nucleon-nucleon
interactions are taken at high energies \cite{landolt}. However, when 
studying antikaon production from heavy-ion collisions at energies 
per nucleon below the  nucleon-nucleon threshold,
cross sections slightly 
above the elementary threshold $\sqrt{s_0}=2m_N+2m_K$ are needed \cite{gqli}.

The particular interest in antikaon production in nucleus-nucleus collisions
at subthreshold energies arises from the expectation that it offers the
possibility to study the in-medium properties of an antikaon at high
baryon density. Indeed, in Ref. \cite{gqli} it has been shown within a  
relativistic transport model \cite{ko,blattel}, that the antikaon yield 
can increase by an order of magnitude if its mass is allowed to decrease 
in dense matter. The experimental data on subthreshold $K^-$-production 
from heavy-ion collisions at GSI \cite{schr,senger} seem to support such 
a scenario.

In fact, according to studies based on effective chiral Lagrangians 
the antikaon mass should be  reduced significantly in the medium due 
to the attractive scalar and vector potentials \cite{kaplan,brown,lee,wass}. 
A strong attractive antikaon potential is also consistent with the kaonic 
atom data \cite{fried}. The extrapolation of this attractive potential 
to high baryon density has led to the suggestion of a possible kaon 
condensate in neutron stars \cite{page}, which then might even lead 
to the formation of mini black holes in galaxies \cite{bethe}.

It is thus of great interest to learn from heavy-ion collisions at
subthreshold energies about the antikaon properties at high densities.
Although the relativistic transport model has been quite successful
in explaining the experimental data on subthreshold 
$K^+$ \cite{fang,lang,cassinga} and antiproton \cite{cassinga,teis,antip} 
production, the results obtained in Ref. \cite{gqli} for subthreshold 
antikaon production depend on the input elementary antikaon production 
cross section from a nucleon-nucleon interaction as well as on the 
treatment of antikaon absorption by nucleons. In Ref. \cite{gqli}
the elementary antikaon production cross section in the parameterization 
introduced by Zwermann and Sch\"urmann \cite{zwermann} (based on data 
from high energy proton-proton scattering) has been used.  The accuracy 
of this parameterization near threshold is not known. Preliminary data on 
$K^-$-production from proton-proton reactions near threshold at 
COSY \cite{cosy} seem to indicate a much smaller cross section
than the one suggested by  Ref. \cite{zwermann}. In order to extract more 
reliable information on antikaon in-medium properties from heavy-ion 
collisions one thus requires a good knowledge on the antikaon production 
cross section from the nucleon-nucleon interaction near threshold.

In the present paper we propose a meson exchange model to evaluate the
antikaon production cross section from a nucleon-nucleon interaction. 
A similar approach (based on Ref.~\cite{yao}) has also been
used to evaluate the kaon production cross section from a nucleon-nucleon 
interaction in Ref.~\cite{texas}. It has been found that this model gives a more
reliable cross section near threshold than the parameterization based 
on $K^+$-data from high energies~\cite{randrup,schur}.

\section{Pion exchange}

The Feynman diagram for antikaon production from a nucleon-nucleon
interaction $NN \to NNK {\bar K}$ in the one-pion exchange model is 
shown in Fig. \ref{fi1}a). Following the derivation in 
Refs. \cite{yao,texas}, the isospin-averaged cross section for this 
process can be written as
\begin{eqnarray}\label{OME}
\bar{\sigma} (NN \to NNK {\bar K}; \sqrt{s}) =C
\frac {m_N^2} {4{\pi}^2 p_i^2 s}
\int _{W_{min}}^{W_{max}}dW \ W^2 k  \nonumber \\
\frac {f^2_{\pi NN} } {{m_\pi}^2} \int _{t_-}^{t_+}dt t F_{\pi}^4(t)
\frac{1}{(t-m_\pi^2)^2} \ \bar{\sigma} (\pi N \to NK{\bar K}; W, t),
\end{eqnarray}
where $\sqrt{s}$ is the center-of-mass energy of the colliding nucleons.  
The energy of the pion-nucleon system in their center-of-mass is given 
by $W$, with values between
\begin{equation}
W_{min}=2m_K + m_N,~~~~~{\rm and}~~~~~W_{max}= \sqrt{s}-m_N,
\end{equation}
where $m_N$ and $m_K$ denote the masses of the nucleon and kaon, 
respectively.  The squared four-momentum transfer from the initial to the 
final nucleon is denoted by $t$ and varies between the two limits,
\begin{equation}
t_{\pm} = 2m_N^2-2E_iE_f \pm 2p_ip_f,
\end{equation}
where $E_i$, $p_i$ are the energy and momentum of the initial nucleons 
in the center-of-mass frame, while $E_f$, $p_f$ are those for the final 
nucleons. They are related to $\sqrt{s}$ by
\begin{eqnarray}
p_i^2  = {\lambda}(s, m_N^2,m_N^2)~~~~~{\rm and}
~~~~~p_f^2  = {\lambda}(s, W^2, m_N^2),
\end{eqnarray}  
with
\begin{equation}
\lambda (x, y, z) = \frac {(x-y-z)^2-4yz} {4x}.
\end{equation}
Similarly, the three-momentum $k$ of the exchanged pion is related to 
$W$ and its mass $m_\pi$ by $k^2 = {\lambda}(W^2, m_N^2, {m_\pi}^2)$.
The pseudovector pion-nucleon coupling is denoted by $f_{\pi NN}$
and has a value of $\approx 1$.
To take into account the off-shell nature of the exchanged pion,
we introduce (as in Ref.~\cite{laget}) at the $\pi NK{\bar K}N$
vertex a pion form factor similar to that in the
$\pi NN$ vertex,
\begin{equation}\label{cutoff}
F_{\pi}(t) = \frac {{\Lambda}^2 - {m_\pi}^2}
{ {\Lambda}^2 - t},
\end{equation}
with ${\Lambda} $ denoting the cutoff parameter.

\section{$\pi N \to N K\bar K$}

\begin{figure}
{\psfig{figure=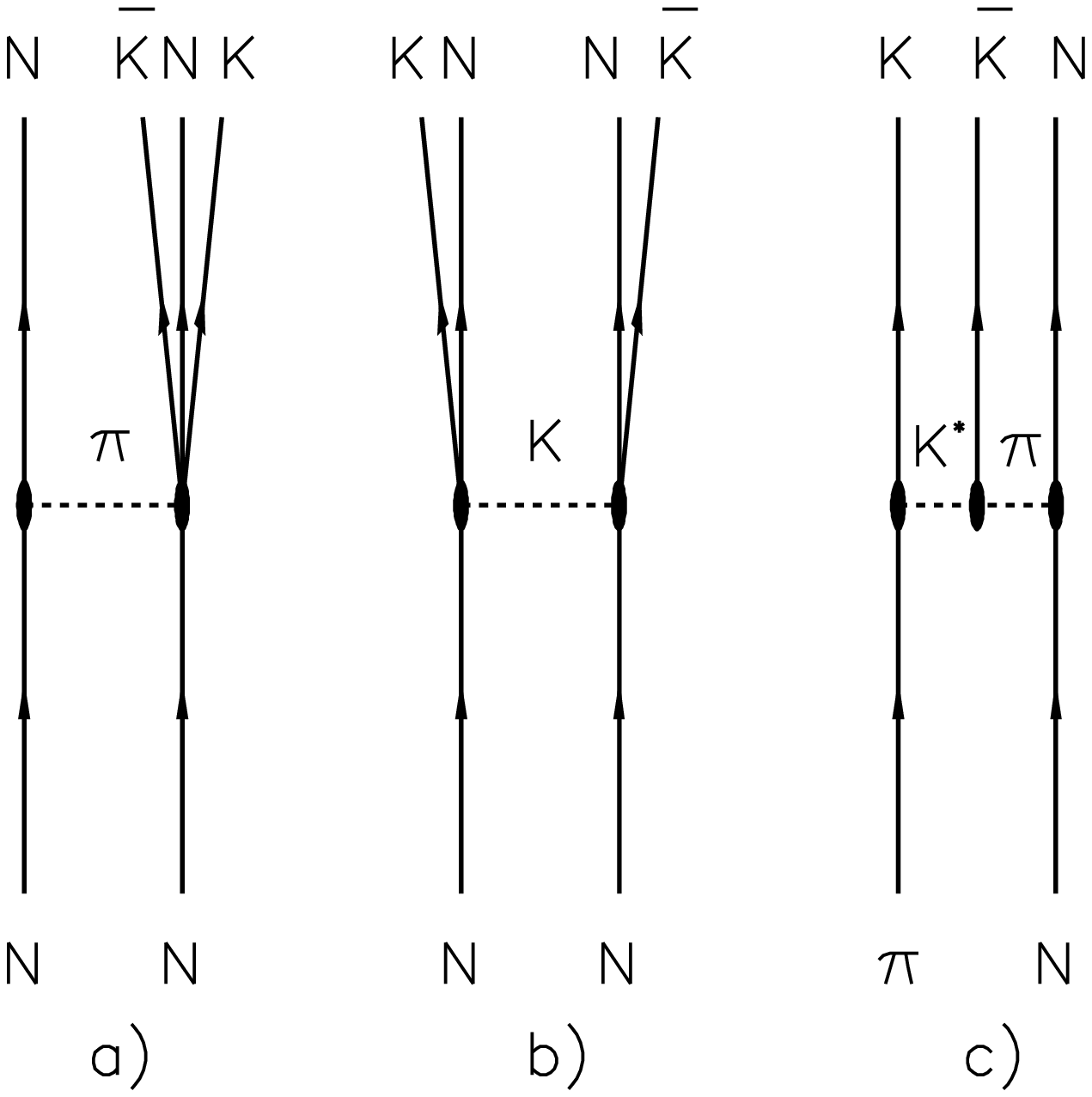,width=14cm,height=10cm}}
\caption{\label{fi1}Feynman diagrams for the reaction
$NN{\to} NNK\bar K$ (a,b) and  $\pi N{\to} NK\bar K$ (c).}
\end{figure}

In Eq. (\ref{OME}), ${\bar\sigma}(\pi N {\to} NK {\bar K} ; W, t)$
is the isospin-averaged on-shell ${\bar K}$ production cross
section from a $\pi N$ interaction. The coefficient $C$ in Eq. (\ref{OME})
depends on how different isospin channels for this reaction are related.  
One possible model for this reaction is shown in Fig. \ref{fi1}c), in
which the pion annihilates with a virtual pion from the nucleon to 
produce the $K\bar K$ pair via the exchange of a $K^*$-resonance. For 
on-shell pions, this process has been shown to give a reasonable 
description of $K\bar K$ production from pion-pion interactions \cite{xia}. 
For simplicity, we neglect in the present work the off-shell effects by 
dropping the $t$-dependence in ${\bar\sigma}(\pi N {\to} NK {\bar K};W, t)$
and using thus the empirical on-shell cross section.

\begin{figure}
{\psfig{figure=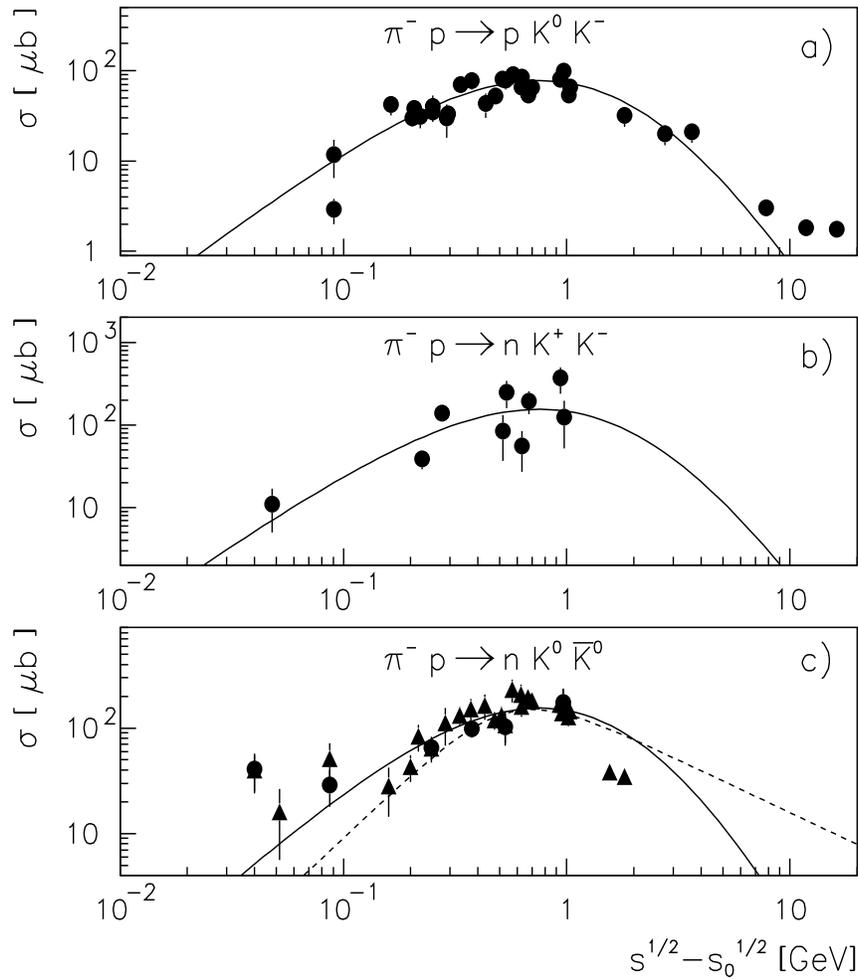,width=14cm,height=15cm}}
\caption{\label{fi3}Cross sections for the reactions:
(a) ${\pi}^-p \to pK^0K^-$ ,
(b) ${\pi}^-p \to n K^+K^-$, and
(c) ${\pi}^-p \to n K^0{\bar K^0}$.
The solid lines show our parameterization
(Eqs. (\protect\ref{par0}) and (\protect\ref{iso})), while
the experimental data are from Ref. \protect\cite{landolt}.
The dashed line is the parameterization (Eq. (\protect\ref{par1}))
from Ref. \protect\cite{paryev}.}
\end{figure}

\begin{figure}
{\psfig{figure=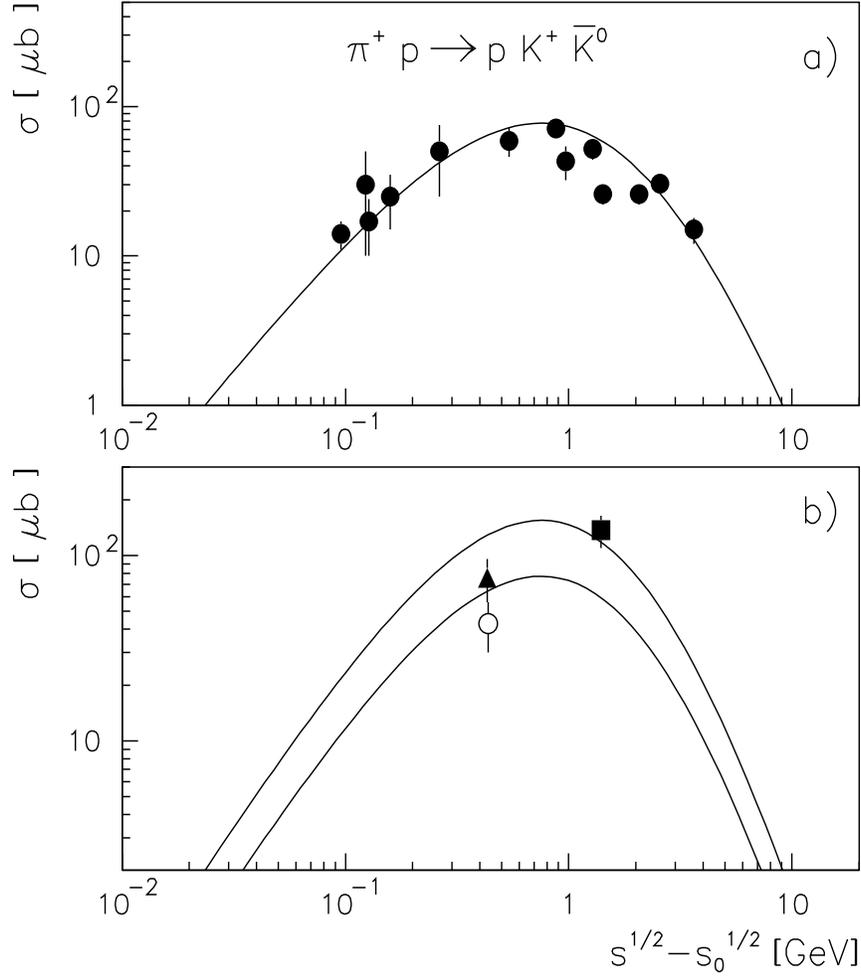,width=14cm,height=15cm}}
\caption{\label{fi4}Same as Fig. 2 for the reactions:
(a) ${\pi}^+p \to pK^+{\bar K^0}$ and
(b) ${\pi}^+n \to nK^+K^-$ (square),
${\pi}^+p \to pK^0 {\bar K^0} $ (triangle), and
${\pi}^+n \to nK^+{\bar K^0}$ (open circle).
The lower solid line in (b) is our result for the reactions
${\pi}^+p \to pK^0 {\bar K^0} $ and ${\pi}^+n \to nK^+{\bar K^0}$,
while the upper solid line in (b) is that for the reaction
${\pi}^+n \to nK^+K^-$.}
\end{figure}

Experimental data on antikaon production from pion-nucleon interactions
are available for the following reactions \cite{landolt}:
$\pi^-p{\to}pK^0K^-$, $\pi^-p{\to}nK^+K^-$, $\pi^-p{\to}nK^0{\bar K^0}$,
$\pi^+p{\to}pK^+{\bar K^0}$, $\pi^+n{\to}pK^+K^-$,
$\pi^+n{\to}pK^0{\bar K^0}$,
and $\pi^+n{\to}nK^+{\bar K^0}$ (cf. Figs. \ref{fi3} and \ref{fi4}).
We first parameterize the experimental $\pi^-p{\to} pK^0K^-$ cross section
by the expression,
\begin{equation}\label{par0}
\sigma ({\pi^-} p{\to} pK^0K^-) = 1.121
{\left( 1- \frac {s_0} {s} \right)}^{1.86}
{\left( \frac {s_0} {s} \right)}^2  \ \ [mb],
\end{equation}
where $\sqrt{s}$ is the invariant mass of the $\pi N$ system
and $\sqrt{s_0}=m_N+2m_K$. The results of the fit are shown
in Fig. \ref{fi3}a) by the solid curve in comparison to the data 
from Ref. \cite{landolt}.  Instead of fitting the other cross sections 
separately, we explore the isospin symmetry of the Feynman diagram 
shown in Fig. \ref{fi1}c), leading to the following relations among
the cross sections:
\begin{eqnarray}\label{iso}
& 2 \sigma ({\pi}^+ p {\to} p K^+ {\bar K^0}) =
2 \sigma ({\pi}^+ n {\to} n K^+ {\bar K^0})= &
\sigma ({\pi}^+ n {\to} p K^+ K^-) =
\nonumber  \\
& \sigma ({\pi}^+ n {\to} p K^0 {\bar K^0}) =
\sigma ({\pi}^0 p {\to} n K^+ {\bar K^0})= &
4 \sigma ({\pi}^0 p {\to} p K^+ K^- ) =
\nonumber  \\
& 4 \sigma ({\pi}^0 p {\to} p K^0 {\bar K^0} )=
\sigma ({\pi}^0 n {\to} p K^0  K^-) = &
4 \sigma ({\pi}^0 n {\to} n K^+ K^- ) =
\nonumber \\
& 4 \sigma ({\pi}^0 n {\to} n  K^0 {\bar K^0} )=
2 \sigma ({\pi}^- p {\to} p K^0 K^- )  = &
\sigma ({\pi}^- p {\to} n K^+ K^- ) =
\nonumber \\
&\sigma ({\pi}^- p {\to} n K^0 {\bar K^0} )=
2 \sigma ({\pi}^- n {\to} n K^0  K^-).&
\end{eqnarray}
All cross sections are thus related to
$\sigma (\pi^-p{\to} pK^0K^-)$; they are shown in
Figs. \ref{fi3}b), \ref{fi3}c), and \ref{fi4})
by the solid curves, and it is seen that they agree quite well with the data.
We, therefore, believe that the isospin relations are properly treated
via the diagram in Fig. \ref{fi1}c) and  that $K^*$-resonance
exchange  is the dominant mechanism.  For comparison, we also show
in Fig. \ref{fi3}b) (dashed curve) the parameterization introduced
in Ref. \cite{paryev}, i.e.,
\begin{equation}\label{par1}
\sigma ({\pi}^- p {\to} n K^0 {\bar K^0}) =
\frac {0.158(\sqrt{s}-\sqrt{s_0})^2} {0.1735 +(\sqrt{s}-\sqrt{s_0})^3} 
\ \ [mb].
\end{equation}
We find that our parameterization fits the data better near threshold 
than that of Ref. \cite{paryev}.

>From Eq.(\ref{iso}) we can also derive the
isospin-averaged cross section for $\pi N{\to} NK\bar K$, which then is
given by
\begin{equation}
{\bar\sigma} (\pi N{\to} N K{\bar K}) = 3\sigma(\pi^-p{\to} pK^0K^-).
\end{equation}
This cross section is useful for transport models that do not explicitly 
include the isospin degree of freedom.

\begin{figure}
{\psfig{figure=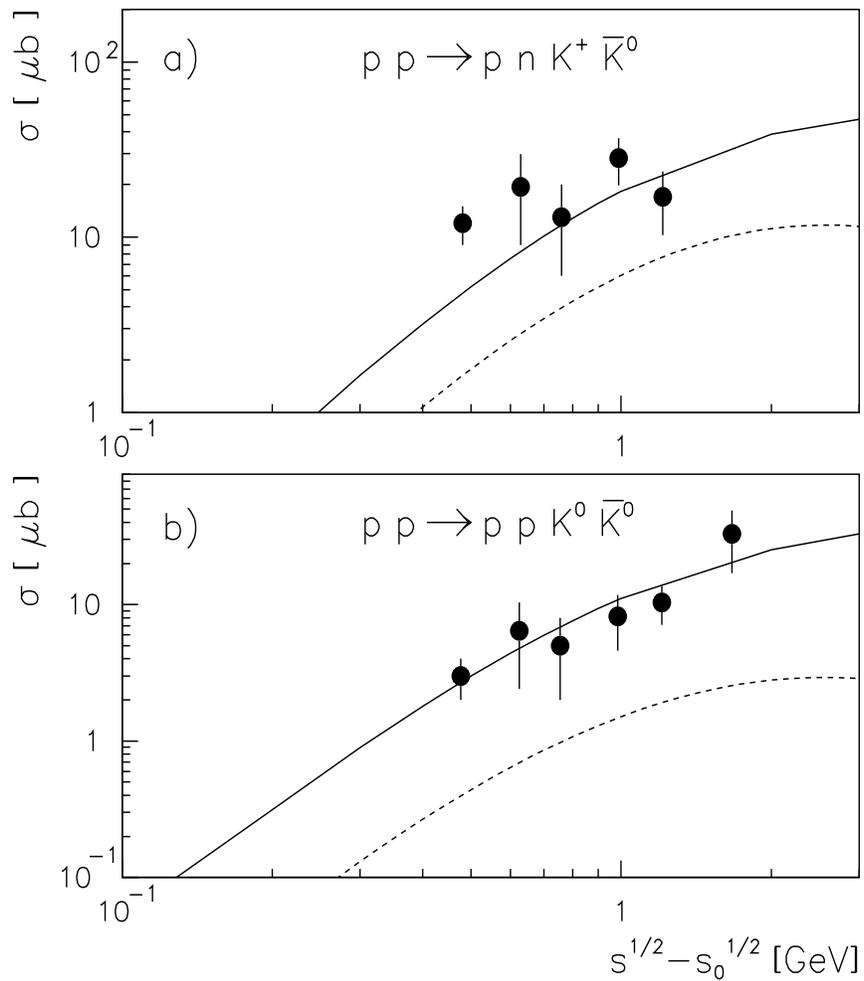,width=14cm,height=15cm}}
\caption{\label{fi5}Cross sections for the reactions:
(a) $pp \to pnK^+ {\bar K^0} $ and
(b) $pp \to ppK^0 {\bar K^0} $.
The dashed lines show our results calculated with the one-pion exchange
model according to Eqs. (\protect\ref{OME}) and (11), while
the solid lines are the total contribution from both pion and kaon exchange.
The experimental data are from Ref. \protect\cite{landolt}.}
\end{figure}

With the model shown in Fig. \ref{fi1}c) for the reaction
$\pi N{\to} NK\bar K$, the coefficient $C$ in Eq. (\ref{OME}) has a value of
7/8. Now substituting the isospin-averaged $\pi N{\to} NK\bar K$ cross section
in Eq. (\ref{OME}), we can evaluate the isospin-averaged cross section for 
antikaon production from nucleon-nucleon interactions. Experimental data 
are available for the reactions $pp {\to} pnK^+ {\bar K^0} $ and
$pp {\to} ppK^0 {\bar K^0}$ \cite{landolt} at high energies. From isospin
symmetry, they can be related to the isospin-averaged cross section by
\begin{equation}
\sigma (pp{\to} pnK^+{\bar K^0})=4\sigma (pp{\to} pp K^0{\bar K^0})
=\frac{16}{21}{\bar \sigma}(NN{\to} NNK\bar K).
\end{equation}
We use a monopole cutoff $\Lambda$=1.2 GeV in Eq.(\ref{cutoff}),
which is taken
from the analysis of $K^+$-meson production \cite{texas}.
Our results calculated with the one-pion exchange model
according to Eq.(\ref{OME}) are shown
in Fig. \ref{fi5} by the dashed lines and are seen to underestimate the
experimental data by about a factor 3-4.

\section{Kaon exchange}

Kaon exchange also contributes to antikaon production from a 
nucleon-nucleon interaction as shown in 
Fig.\ref{fi1}b. Neglecting interferences,
the cross section can be expressed in
terms of the product of elastic $KN$ and $\bar KN$ cross sections
\cite{yao,chuk}. Both isospin ( I=1 and I=0) amplitudes contribute to
the latter reactions. If their interference is neglected,
we can relate the I=1 and I=0 cross sections to the empirical ones
\cite{cugnon}, i.e.
\begin{eqnarray}
\sigma_1(KN{\to}KN)&=&\sigma(K^+p{\to}K^+p) \nonumber \\
\sigma_0(KN{\to}KN)&=&\sigma(K^+n{\to}K^+n)+\sigma
(K^+n{\to}K^0p)  \nonumber \\
&-&\sigma(K^+p{\to}K^+p)
\end{eqnarray}
and 
\begin{eqnarray}
\sigma_1({\bar K}N{\to}{\bar K}N)&=&\sigma(K^-n{\to}K^-n)\nonumber\\
\sigma_0({\bar K}N{\to}{\bar K}N)&=&\sigma(K^-p{\to}K^-p)+\sigma
(K^-p{\to}K^0n) \nonumber \\
&-&\sigma(K^-n{\to}K^-n).
\end{eqnarray}
The cross sections on the right hand side of Eqs. (12) and (13) are
available empirically and used in the  parameterizations of
Ref. \cite{cugnon}. We note that the results for $\sigma_0(KN)$ and
$\sigma_1(KN)$ are in reasonable agreement with the calculations
on $KN$ elastic cross sections performed recently by
Hoffmann et al.~\cite{hoffmann}.

In terms of the above cross sections we have
\begin{eqnarray}\label{OKE}
&&\sigma(pp{\to}ppK^0\bar{K^0};\sqrt{s})=
\frac {1} {2{\pi}^3 p_i^2 s}
\int _{W_{min}}^{W_{max}} dW \ W^2  k_1 [\sigma_1(KN{\to}KN;W) \nonumber  \\
&&+\sigma_0(KN{\to}KN;W)] \ \int _{U_{min}}^{U_{max}}dU \ U^2 k_2
\sigma_1(\bar{K}N{\to}\bar{K}N;U) \nonumber \\
&&\times \int _{t_-}^{t_+} dt \ F_K^4(t) \frac {1} {(t-m_K^2)^2}.
\end{eqnarray}
In Eq. (\ref{OKE}) $\sqrt{s}$ is again the center-of-mass energy of the
colliding nucleons,
$W$ is the energy of the kaon-nucleon system in their center-of-mass,
and $U$ is that of the antikaon-nucleon system. The allowed values
for $W$ and $U$ are given by 
\begin{eqnarray}
W_{min}&=&m_K + m_N,~~~~~~~~~~W_{max}= \sqrt{s}-m_N-m_K,
\nonumber \\
U_{min}&=&m_K + m_N,~~~~~{\rm and}~~~~~U_{max}=\sqrt{s}-W.
\end{eqnarray}
The three-momenta $k_1$ and $k_2$ of the exchange kaon in the
$KN$ and $\bar KN$ system are related to $W$ and $U$ by 
\begin{equation}
k^2_1 = {\lambda}(W^2, m_N^2, m_K^2) \ \ \ {\rm and} \ \ \
k^2_2 = {\lambda}(U^2, m_N^2, m_K^2).
\end{equation}
The squared four-momentum transfer from the initial nucleon to the 
final $KN$ system is denoted by $t$ and varies between the two limits,
\begin{equation}
t_{\pm} = W^2+m_N^2-2E_iE_f \pm 2p_ip_f,
\end{equation}
where $E_i$, $p_i$ are the energy and momentum of the initial nucleons 
in the center-of-mass frame defined as before,
while $E_f$, $p_f$ are those for the final $KN$ system
and are given by 
\begin{equation}
E_f=\frac {(s+W^2-U^2)} {2\sqrt{s}} \ \ \ {\rm and} \ \ \
p^2_f = {\lambda}(s, W^2, U^2).
\end{equation}
We account for the off-shellness of the exchanged kaon at
both $KN$ and ${\bar K}N$ vertices by introducing a
monopole form factor with the cutoff parameter
$\Lambda $=1.0 GeV, which is close to that used in Ref.~\cite{texas}.

Similarly, one can show that 
\begin{eqnarray}
&&\sigma(pp{\to}pnK^+\bar{K^0};\sqrt{s})=
\frac {1}{2{\pi}^3 p_i^2 s} \nonumber \\
&&\left[ \int _{W_{min}}^{W_{max}}dW \ W^2 k_1 \sigma_1(KN{\to}KN;W)
\int _{U_{min}}^{U_{max}} dU \ U^2 k_2
\sigma_1(\bar{K}N{\to}\bar{K}N;U) \right. \nonumber\\
&&+\frac{1}{2} \int _{W_{min}}^{W_{max}} dW \ W^2 k_1
\sigma_1(KN{\to}KN;W)
\int _{U_{min}}^{U_{max}} dU \ U^2 k_2
\sigma_0(\bar KN{\to}\bar KN;U) \nonumber \\
&&\left. +\frac{1}{2} \int _{W_{min}}^{W_{max}} dW \ W^2 k_1
\sigma_0(KN{\to}KN;W)
\int _{U_{min}}^{U_{max}} dU \ U^2 k_2
\sigma_1(\bar KN{\to}\bar KN;U) \right] \nonumber \\
&&\times \int _{t_-}^{t_+}dt \ F_K^4(t) \ \frac {1} {(t-m_K^2)^2}.
\end{eqnarray}

The isospin-averaged antikaon production
cross section due to kaon exchange is then given by
\begin{eqnarray}
&&\bar{\sigma}(NN{\to} NNK\bar K;\sqrt{s}) =
\frac {1} {{\pi}^3 p_i^2 s} \nonumber \\
&&\left[ \int _{W_{min}}^{W_{max}} dW \ W^2 k_1
\sigma_1(KN{\to} KN; W)
\int _{U_{min}}^{U_{max}} dU \ U^2 k_2
\sigma_1(\bar{K}N {\to} \bar{K}N; U)
\right. \nonumber\\
&&+\frac{1}{2}
\int _{W_{min}}^{W_{max}} dW \ W^2 k_1
\sigma_1(KN{\to} KN;W)
\int _{U_{min}}^{U_{max}} dU \ U^2 k_2
\sigma_0(\bar KN{\to}\bar KN;U) \nonumber \\
&&+\frac{1}{2}
\int _{W_{min}}^{W_{max}} dW \ W^2 k_1
\sigma_0(KN{\to} KN;W)
\int _{U_{min}}^{U_{max}} dU \ U^2 k_2
\sigma_1(\bar KN{\to}\bar KN;U) \nonumber \\
&&+\frac{1}{8}
\left. \int _{W_{min}}^{W_{max}} dW \ W^2 k_1
\sigma_0(KN{\to} KN;W)
\int _{U_{min}}^{U_{max}} dU \ U^2 k_2
\sigma_0(\bar KN{\to}\bar KN;U)  \right]. \nonumber \\
&&\times \int _{t_-}^{t_+} dt \ F_K^4(t) \frac {1} {(t-m_K^2)^2}.
\end{eqnarray}

The contribution to antikaon
production from the kaon exchange is found to be more important
than that from the pion exchange. 
The total contributions from both ( pion and kaon)  exchanges are shown in
Fig.~\ref{fi5} by the solid lines and are seen to reproduce reasonably
well the experimental data.

Furthermore, our numerical results for the isospin averaged antikaon
production cross section from the nucleon-nucleon interaction
calculated with the pion and kaon exchange model can be
parameterized by 
\begin{equation}
{\bar\sigma}(NN{\to} NNK\bar K, \sqrt{s})=
 \ 0.3 {\left( 1-\frac{s_0}{s} \right)}^{3.0}
{\left( \frac{s_0}{s} \right)}^{0.8} \ \ [mb],
\end{equation}
where $\sqrt{s_0}=2m_N+2m_K$. Again, this will be useful for transport 
model calculations without explicit treatment of the isospin degrees
of freedom.

\section{Discussion}

\begin{figure}
{\psfig{figure=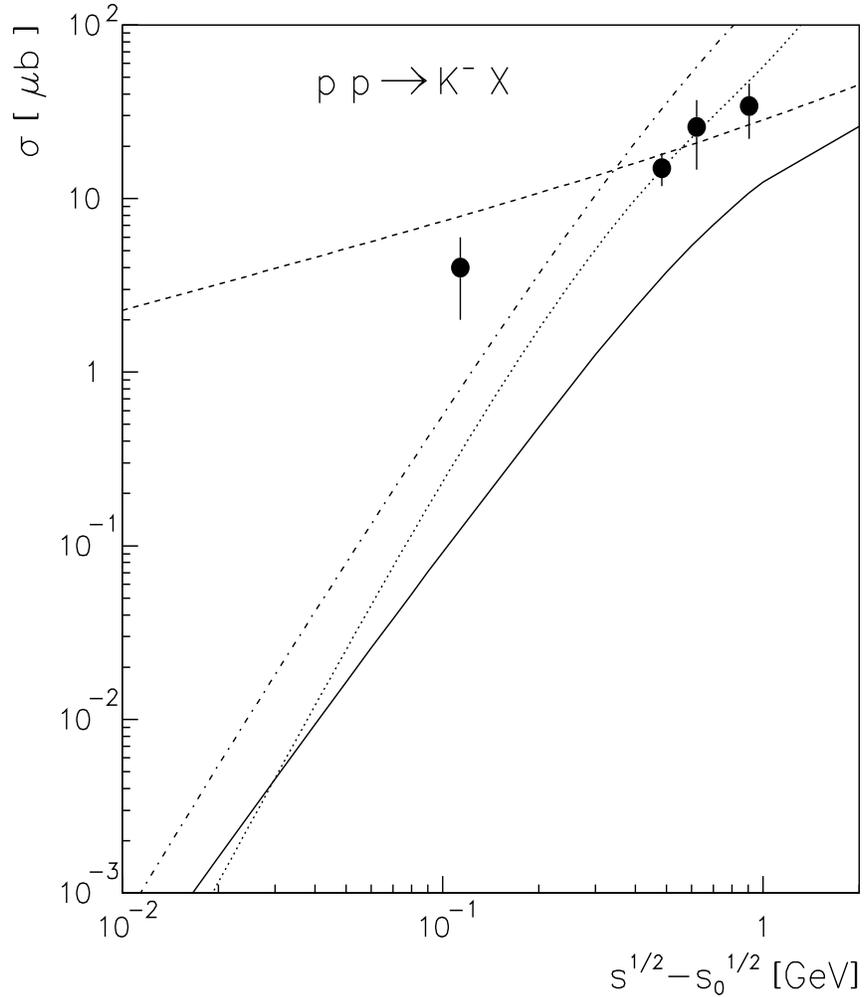,width=14cm,height=15cm}}
\caption{\label{fi6}$K^-$ inclusive production cross section from 
proton-proton interactions in comparison to the experimental data 
from Ref. \protect\cite{landolt}. The solid line shows our calculation
for the exclusive reaction $pp \to p p K^+K^-$;
the dashed line represents the
parameterization from Ref. \protect\cite{zwermann}, the dotted line
is the parameterization from Ref. \protect\cite{paryev1}, and the 
dash-dotted line from the statistical ROC model of  
Refs. \protect\cite{sibirtsev1,muller}.}
\end{figure}

We show in Fig. \ref{fi6} the available parameterizations for
the inclusive $K^-$ production from $pp$ collisions together with the 
experimental data \cite{landolt}.  We note, that our result (solid
line) includes the contributions from pion and kaon exchanges, but
accounts only for the exclusive reaction $pp {\to} pp K^+ K^-$.
Obviously we underestimate the data
at higher energies since here processes with one
$(\sqrt{s}-\sqrt{s_0}>>m_{\pi})$ and more pions in the
final state  also contribute to the inclusive cross section.

For comparison, the dash-dotted line in Fig. \ref{fi6}
shows the results from Ref. \cite{sibirtsev1}
calculated with the statistical quark ROC model \cite{muller}, while the 
dotted line is the parameterization from Ref. \cite{paryev1} based on 
phase space considerations.
The dashed line in Fig. \ref{fi6} is the parameterization
from \cite{zwermann} and differs drastically from our results at low
energies since it tries to fit the data point at about 0.1 GeV above 
threshold from Ref. \cite{reed},
which does not follow the systematics implied 
by phase space considerations. It is thus very important to have 
experimental data at low energies to determine if our model is
appropriate. In this respect, experiments currently being carried 
out at COSY (at 6 MeV above threshold)~\cite{ki} are extremely useful.

\section{Summary}

In summary, we have introduced a meson exchange model for antikaon
production from nucleon-nucleon interactions. This model allows us 
to express the cross section in terms of the off-shell  production
amplitudes from the pion-nucleon and kaon-nucleon interactions.
Approximating the off-shell
amplitudes by the measured on-shell amplitudes and neglecting
their interferences, the
$NN{\to} NNK\bar K$ cross section can be easily calculated at
low energies, where the cross section is needed for studying antikaon 
production in proton-nucleus and nucleus-nucleus collisions at subthreshold
energies.

Comparing with earlier parameterizations, our predictions give 
a smaller  cross section near threshold, roughly in line
with the phase space considerations from Ref. \cite{paryev1}. 
It is thus
important to have experimental data from  nucleon-nucleon interactions 
at low energies to verify our predictions.

As a byproduct of our study we have found that the available data
for the antikaon production cross section from the different isospin 
channels in the reaction $\pi N{\to} NK\bar K$ are consistent with a
model in which the pion interacts with a virtual pion from the nucleon 
through the exchange of a $K^*$-resonance.  A simple parameterization
is introduced and is able to account for all available data for this
reaction.

We note that the study of  antikaon production from subthreshold 
heavy-ion collisions requires the knowledge of both nucleon-nucleon
and pion-nucleon production cross sections.  In Ref. \cite{gqli}, the 
parameterization of Zwermann and  Sch\"urmann \cite{zwermann}, which 
is  a few orders of magnitude larger than ours near threshold,
has been used; yet the results have indicated that a strong attractive 
antikaon potential is needed in order to explain the measured cross 
section from Ref. \cite{schr}.  With a smaller elementary 
cross section according to our prediction, the
antikaon medium effects should be even more pronounced.

\section*{Acknowledgement}

The authors like to acknowledge valuable discussions with U. Mosel 
throughout this study.   C.M.K. was also supported in part by the 
National Science Foundation under Grant No. PHY-9509266 and the Alexander 
von Humboldt Foundation.

\newpage

\end{document}